\documentclass[3p]{elsarticle}

\usepackage{graphicx,amssymb}

\usepackage[english]{babel}
\usepackage{xcolor}
\definecolor{hypercolor}{rgb}{0,0.2,0.7}

\usepackage[breaklinks,final]{hyperref}
\hypersetup{
  colorlinks,
  linkcolor=hypercolor,
  urlcolor=hypercolor,
  citecolor=hypercolor,
  bookmarksnumbered,
  bookmarksdepth=2,
  pdfauthor={Ivan Zhogin},
  pdftitle={One more fitting (D=5) of Supernova redshifts}}

\def\d{\delta}               
                    \def\l{\lambda}
\def\m{\mu}             \def\n{\nu}              
                       
      \def\om{\omega}
\def\G{\Gamma}          \def\D{\Delta}           \def\L{\Lambda}
          \def\Om{\Omega}

\def\be{\begin{equation}}            \def\ee{\end{equation}}
\def\ba#1{\begin{array}{#1}}         \def\ea{\end{array}}

\def\fr#1#2{\textstyle\frac{#1}{#2}}
 
\def\cL{\mathcal{L}}

\journal{arXiv}
\begin{document}
\begin{frontmatter}
\title{One more fitting ($D{=}5$) of Supernova redshifts}
\author{I.\,L. Zhogin\fnref{fn1}}

\address{ \  \ } %
\fntext[fn1]{zhogin@mail.ru}

\begin{flushright}
{\normalsize \emph{
 In memory of Dr. N.\,G. Gavrilov}}
\end{flushright}

\begin{abstract}
Supernova Ia redshifts are fitted with a simple $5D$ model: the
galaxies are assumed to be enclosed in a giant $S^3$-spherical
shell of significant thickness, which expands (ultra)relativistically in a (1+4)$D$ Minkowski space. This model, as compared with the kinematic
(1+3)$D$ Milne cosmological model (which was reinvented by Prof Farley), 
goes in line with the Copernican
principle: any galaxy observes the same isotropic distribution
of distant galaxies and supernovae, as well as the same Hubble plot of
SN Ia distance modulus $\mu$ vs redshift $z$. A good fit is obtained
(no free parameters); it coincides with the Milne model (empty model)
at low $z$,
while shows some more luminosity at high $z$, leading to 1\%
decrease in the true distance modulus (and 50\% increase in
luminosity) at $z\sim2$.

The model proposed can be also interpreted as a sort of FLRW-model
with the scale factor $a(t)=t/t_0$; this could hardly be a solution
of general relativity (without inventing some super-dark energy
with $w=-1/3$, a sort of dark curvature); 
5$D$ GR is also unsuitable -- it has no
longitudinal polarization. However, there still exists the
other theory (with $D=5$ and no singularities in solutions), the
other game in the town, which seems to be able to do the job.
\end{abstract}
\end{frontmatter}
 \noindent
1. The supernovae of type Ia (SNe Ia) as a rule can
serve as standard candles. To fit the SNeIa data
 (redshifts \emph{vs} luminosities), the
general relativity theory (GR) requires dark energy, dark matter, and a
few fitting parameters. 
 Recently the Hubble diagram was excellently fitted
 in framework of a very simple model based on 
special relativity \cite{far} (Farley, 2009). The galaxies are
assumed to recede with constant velocities in the usual Minkowski
space, $D=1+3$. Taking into account all relativistic effects (time
dilation and so on), one can find the SN luminosity   (at the
distance $r$, when the light was emitted) as function of redshift $z$
\cite{far}:
\[  L=L_0
\frac{d_*^2}{c^2t_0^2}\,\frac{(1+\beta)^2}{\beta^2(1+z)^4}, \
\mbox{where } \, (1+z)^2=\frac{1+\beta}{1-\beta}, \ \,
\beta=\frac{z(2+z)}{2+2z+z^2}=\frac{r}{ct_0-r}\, ;
\]
 $d_*=10$ pc, the velocity is $\beta c$,
 $t_0$ is the time from the big bang; the  Hubble
 constant (as accepted in \cite{hick}) is
 \[ H_0=t_0^{-1}\approx65\,\rm{\,km\,s}^{-1\,}\rm{Mpc}^{-1}\,.\]

In fact, this model is just the old special-relativistic model suggested by
Milne \cite{milne}; see also \cite{dm} 
where Milne's model, its lack of gravity, is  
explained with antigravity of antimatter.\footnote{Following this logic, 
the photons which coincide with their antiparticles should have no gravity, but 
this is not the case.} 

 In magnitudes, the \emph{true distance modulus}
  of a SN Ia should be
 (everywhere $\log$ means $\log_{10}$) \cite{far}
 \be \label{one}  \mu(z)=
 -2.5\log(L/L_*)=\mu_0+5\log[\beta/(1-\beta)]
 =\mu_0+5\log(z+z^2/2)\,.
 \ee

Below I consider another model which gives the next solution:
 \be \label{two}
 \mu(z)=\mu_0+5\log[(1+z)\,\ln(1+z)]=
 \mu_0+5\log[z+z^2/2-z^3/6+\cdots +(-)^n z^n/(n^2-n)+\cdots] \,;
 \ee
 in both cases, by definition, $\, \mu(z_*)\equiv 0, \
  \mu_0=-5\log(z_*)=-5\log( H_0d_*/c)\approx 43.3\, \
 (z_* =\beta_*\ll 1)$.

 As Wikipedia says, the (true) distance modulus $\mu = m - M$ is the
difference between the apparent magnitude $m$ (corrected for
interstellar absorption) and the absolute magnitude $M$ of an
astronomical object; the last
 is defined as the apparent magnitude of an
object when seen at a distance of 10 parsecs ($d_*$).
To find out more about the issue, it is worth reading the
\emph{Dark Energy Primer} in \cite{detf}, as well as the very
interesting remarks on the status of $\Lambda$CDM-model
\cite{lieu} (Lieu, 2007).

The ``Union'' compilation of  SNe Ia data, which includes more
than 300 SNe after selection cuts, has been given by Kowalski et
al \cite{kow}. Even more vast compilation enriched with low-$z$
SNe has been recently given by Hicken et al \cite{hick}, and I
will use the last data (presented in the Table 1 of \cite{hick};
the last footnote to this table specifies $H_0$), with the
following selection cut imposed on the uncertainties of $\mu$,
$\d \mu$:
 \be\label{cut} 100\, \frac{\d\mu}{\mu}< 0.64+z \,
.\ee

\vspace{-5mm}
\begin{figure}[hbt]
\begin{center}   
\includegraphics*[bb=15 65 265 185,width=110mm]{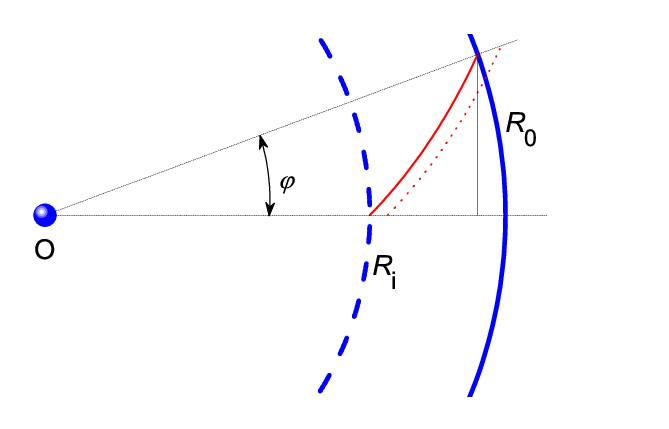}
\caption{Relativistically expanding $S^3$-spherical shell; photons
move along the spirals.}
\end{center}
 \label{fi1}
\end{figure}

\vspace{2.5mm}
 \noindent 2. Let us consider a 5$D$ cosmological
model in which our Universe is a sphere $S^3 $, or more exactly a
spherical shell of some thickness, which expands in a 5$D$
Minkowski space. It is supposed that both matter and light are
kept or confined inside this shell, as in some kind of a wave
guide, and that the speed of expansion, $V$, is close to the unit (put
$c=1 $):
 \[ R=VT \approx T, \ \G=(1-V^2)^{-1}\gg 1\, .\]
Caps letters relate to the privileged reference frame, to the
\emph{special observer} which rests in the center of this sphere. In this
frame, the radial and angle components of the speed of a photon,
 $U_R$ and $U_{\varphi}$,
read (after Lorentz's transformation from a comoving reference
system;  for a comoving observer, the shell thickness is large
enough to neglect the light dispersion):
 \be \label{vv}
 \Om\,\left(%
\begin{array}{c}  1 \\ U_R  \\   U_{\varphi} \\ \end{array}%
\right)  =  \left(%
\begin{array}{ccc}
 \G & \G V & 0 \\  \G V & \G & 0 \\ 0 & 0 & 1 \\ \end{array}%
\right) \,   \left(%
\begin{array}{c}  \om \\ 0  \\ \om \\ \end{array}%
\right); \ \ \textrm{ hence, }  \Om=\G \om, \ U_R=V, \ U_\varphi=\G^{-1}\,.
 \ee
So, the light moves along the spirals in accordance with the next
equations:
 \be\label{spi}
 \frac{d\,R}{d\,T}=V,\ R\frac{d\,\varphi}{d\,T}=\frac1{\G}; \
 R=VT, \ \varphi= \frac1{\G V} \, \ln\frac{T}{T_i}, \
 \frac{R}{R_i} = \frac{T}{T_i} = e^{\G V \varphi}.
\ee
 Figure 1 illustrates this and shows two photon paths for
 a semi-relativistic case, with $\G V\sim 1$.
 It is clear that the time dilation, with respect to the special observer,
 is the same for all comoving (and \emph{resting}, $\varphi$=const, i.e.\ seeing
negligible dipole anisotropy of the CMB) observers, for all "resting" galaxies,
 $\D t = \D T/\G$; so the mutual time dilation concerned with the
 light, or the redshift $z$,
 $1+z=\D t_0/\D t_i$, can be easily derived:
 \be \label{z}
 1+z = \frac{\D T_0}{\D T_i}=\frac{T_0}{T_i} = e^{\G V \varphi}.
\ee
 It means that the angle $\varphi$ is very small,
  $\varphi\sim \G^{-1} \ln(1+z) \ll 1$ (even for $z\sim 10^3$).

 \begin{figure}[th]
\begin{center}
\includegraphics[bb=15 15 325 151,width=145mm]{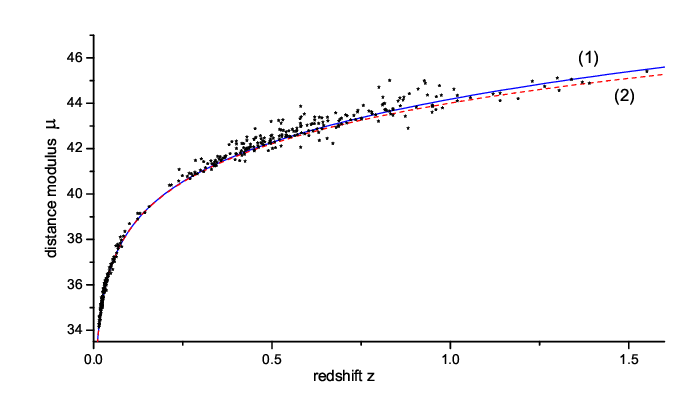}
\caption{Observed SN magnitudes and equations (1) and (2) vs
redshift; $\mu_0=43.3\,$.}
\end{center}  \label{fi2}
\end{figure}
A bunch of photons radiated by some SN at $R_i$ forms at $R_0$ a
sphere of radius [see Eq.\ (\ref{z}) and Fig.~1]
 \[r=R_0 \sin\varphi\approx \G B t_0 (\varphi - \varphi^3/6)
 = t_0 \ln(1+z) [1- \ln^2(1+z) /(6V^2\G^2)];\]
 hence the SN luminosity (i.e. the power flow to a detector
 of unit surface) should look as follows:
 \be \label{lum2}
 L \propto \frac1{(1+z)^2\, \ln^2(1+z)}\, ;
 \ee
here we discard the corrections $\sim \ln^2(1+z)/\G^2$;
 the factor $(1+z)^{-2}$ accounts for reduction of both the
energy of photons and the frequency of their occurrence in a
detector, see also \cite{far}. This equation leads easily to
the plot (\ref{two}); Figures 2 and 3 show the observed data \cite{hick}
and the model curves (\ref{one}), blue, and (\ref{two}), red.

 \begin{figure}[bht]
\begin{center}
\includegraphics[bb=15 15 325 195,width=145mm]{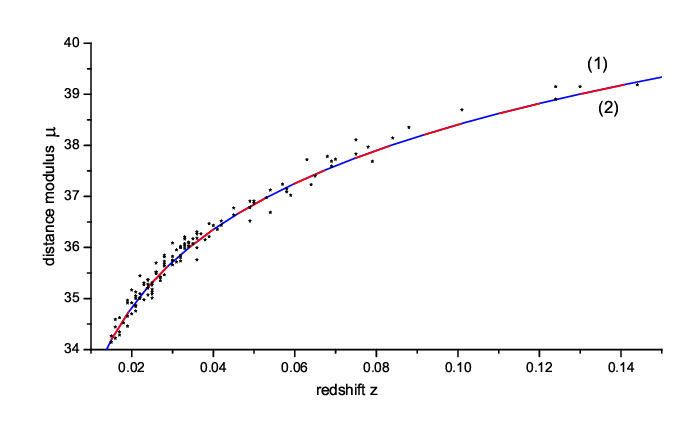}
\caption{The low-$z$ region; curves (1) and (2) look the same.}
\end{center}
 \label{fi3}
\end{figure}

Note that the radius of the Universe is much-much larger 
(with the $\G$ factor) than
the Hubble time $t_0$, and the space curvature is really very small.

The inertial motion of a body with the mass $m$ follows from the Lagrangian:
\[ \cL_* = - m \sqrt{1 - V^2 - V^2 T^2 \varphi^2_{,T} }; \
\,  m V^2 T^2 \varphi_{,T}/\sqrt{1 - V^2 - V^2 T^2 \varphi^2_{,T} } =
\textrm{ const }.
\]
For a comoving resting observer 
($\varphi=\,$const; an ensemble of such observers) with
their time $t$, $ t = T/\sqrt{1-V^2} = T/\G$, the Lagrangian looks as follows
(we use $\dot{\varphi} = d\varphi/dt; \ \psi =\G V \varphi;
 \ x = t \psi = R \varphi$):
\[ \cL = -m \sqrt{1-t^2 \G^2 V^2 \dot{\varphi}^2 } = 
-m \sqrt{1-t^2 \dot{\psi}^2 } =
-m \sqrt{1- ( \dot{x} - x/t)^2; }
\]
\[ m t^2\dot{\psi}/\sqrt{1-t^2\dot{\psi}^2 } =\, \textrm{const}, \]
 this constant is a sort of adiabatic invariant, the radius times the momentum ($p t = p_0 t_0$).
Momentums of both photons and massive moving particles decrease with time, and both make 
work with their centrifugal forces in the moving shell (working thus against the gravity forces; perhaps the zero-point waves are also of no dangerous for cosmological expansion).

Expanding in the series, one can note 
\[ \cL + m \approx m(\dot{x}-x/t)^2/2 + m(\dot{x}-x/t)^4/8
\approx m\dot{x}^2/2 + m \dot{x}^4/8 - 3m\dot{x}^3 x/t - m (x^2/t\dot{)}.
\]

The velocity with respect to the \emph{current} resting observer, $u = t\dot{\psi}$,
defines seemingly the dipole anisotropy of the CMB (for the moving body), and it decreases with time
(a sort of Aristotelean feature, with absolute notion of the rest):
\[  \dot{u} = -(1-u^2)u/t; \, \textrm{while } 
\ddot{x} = u^3/t = (\dot{x} - x/t)^3/t.  \]

The presence of gravity make the situation  more complex.

It  is also interesting to estimate the observed total
$z$-density, $f_{\rm obs}(z)=dM_{\rm obs}/dz$ (that is, baryon
density). Let $\rho(T)$ is the (3$d$ volume) density of matter;
then one can find
 \be\label{dens}
 \D M_{\rm obs} \sim \rho R_i^3 \varphi^2 \D \varphi\,,
 \ \, f_{\rm obs}\propto \frac{\rho(z)\, \ln^2(1+z)}{(1+z)^4}
 \,. \ee
 Different assumptions about $\rho(z)$ can be made and tested (the very
 possibility that our Universe is a non-isolated system
 is very amazing):

 (a) the total number of baryons is a constant; then
 \[ \rho T^3 ={\rm const}\,, \ \, \rho\propto (1+z)^{-3}\, ;\]

the maximum of the function $f_{\rm obs}(z)$ will be at 
$z^*= e^{2/7}-1 \approx 0.331$;

 (b) the number of baryons depends on the 4$d$ volume swept by the
 spherical shell (if it is single):
  some perturbations hit the shell and produce new
 matter (and light as well); in this case (if the rate of
 production does not change)
 \[ \rho T^3 \propto T^4\,, \ \, \rho\propto (1+z)^{-1}\, ;\]

in this case the point of baryon maximum is $z^*= e^{2/5}-1 \approx 0.492$;

  (c) perhaps another simple regime, $\rho\sim \,$const
   or maybe $\rho T^3\sim T$, is also
 reasonable (say, if our shell follows another one).
\vspace{3.5mm}

The similar distributions for SNe and galaxies depend on the star formation 
processes and should be some different.

\noindent 3. It is worth adding a few words about the theory
which in my opinion can support the proposed 5$D$ model.

Although some variants of Riemann-squared modified gravities
(e.g. $R_{\mu\nu}G^{\mu\nu}$-gravity, with equations of forth
order) have a longitudinal polarization related to the Ricci
scalar, these theories are not appropriate \cite{zh1}:
 the polarizations  related to the Weyl tensor
 (and responsible for gravity, tidal forces) are linearly unstable.

Therefore, my main candidate is the theory of frame field,
$h^a{}_\mu$, also known as Absolute Parallelism (AP). AP
benefits from its very high symmetry group which includes both the
global symmetries of Special Relativity (this global subgroup
defines the spacetime signature) and the local symmetries, the
pseudogroup {\it Diff}$(D)$, of General Relativity theory. AP
has the unique variant (no free parameters; with the unique $D$,
$D=5$) where solutions of general position are eternal and free
of arising singularities \cite{zh2}:
 \begin{equation} \label{ue}
 {\bf E}_{a\mu}=L_{a\mu\nu;\nu}- \frac13 (f_{a\mu}
 +L_{a\mu \nu }\Phi _{\nu })=0\, ,
\end{equation}
 where $\, L_{a\mu \nu }=L_{a[\mu \nu]}=
\Lambda_{a\mu \nu }-S_{a\mu \nu }-\frac23 h_{a[\mu }\Phi_{\nu]},
$

\begin{equation}\label{defin}
\Lambda_{a\mu \nu }=2h_{a[\mu,\nu]}, \ S_{\mu \nu
\lambda}=3\Lambda_{[\mu \nu \lambda]}, \ \Phi_\mu=\Lambda_{aa
\mu}, \ f_{\mu\nu}=2\Phi_{[\mu,\nu]}=2\Phi_{[\mu;\nu]}  .
\end{equation}
Coma "," and semicolon ";" denote partial derivative and usual
covariant differentiation with the symmetric Levi-Civita connection,
respectively; our choice is
$\eta_{ab}=\mbox{diag}(-1,1,\ldots,1)$, then
 $g_{\mu\nu}=\eta_{ab} h^a{}_\mu h^b{}_\nu$,  and for any $D$,
$h=\det h^a{}_\mu=\sqrt{-g}$.

One should retain the identities:
 $
 \Lambda_{a[\mu\nu;\lambda]} \equiv 0\,,
  \ \  h_{a\l}\Lambda_{abc;\l}\equiv f_{cb}\,
  (= f_{\m\n}h_c^{\;\m} h_b^{\;\n}), \ f_{[\m\n;\l]}\equiv0
  $.

The equation ${\bf E}_{a\mu;\mu}=0$ gives
 `Maxwell-like equation' (for brevity
 $\eta_{ab}$ and $g^{\m\n}=h_a^{\;\mu} h_a^{\;\nu}$ are omitted
 in contractions):
\begin{equation}\label{max}
(f_{a\mu}
 +L_{a\mu \nu }\Phi _{\nu })_{;\mu}=0, \mbox{ or \ }
 f_{\mu\nu;\nu}=(S_{\mu \nu\l }\Phi _{\l })_{;\n} \ \
(= -\fr1 2 S_{\mu \nu\l }f_{\n\l}, \mbox{ see below}) \, .
\end{equation}
Really (\ref{max}) follows from the symmetric part, because
skewsymmetric one gives just the identity; note also that the
trace part
 becomes irregular (the principal
 derivatives vanish) if $D=4$ (the forbidden number):
 \[2{\bf E}_{[\nu\mu]}=S_{\mu\nu\l;\l}=0, \ {\bf
E}_{[\nu\mu];\nu}\equiv 0; \ \ {\bf E}_{\m\m}={\bf
E}_{a\m}h_b^{\;\m}\eta^{ab} =\fr{4-D}3 \Phi_{\m;\mu}+ (\L^2)=0.
 \]
The system (\ref{ue}) remains compatible under adding
$f_{\m\n}=0$, see (\ref{max});  this is not the case for another
covariant, $S, \Phi$, or Riemannian curvature, which relates to
$\L$ as usually:
\[ R_{a\mu\nu\lambda}=
2h_{a\mu;[\nu;\lambda]}; \ h^a{}_{\mu}h_{a\nu;\lambda}=\frac12
S_{\mu\nu\lambda}-\Lambda_{\lambda\mu\nu}.\]

At first sight this theory does not have a longitudinal
polarization because in linear approximation
\[ R\simeq -2 \Phi_{a,a}\simeq0; \]
 that is, Ricci scalar does not provide a polarization.

However in the case of high symmetry (like spherical symmetry or
plane longitudinal wave, i.e.\ isotropic in the tangential
dimensions), skew-symmetric tensors tend to become zero:
$S_{\mu\nu\lambda}\equiv 0$, and integration of Eqs.~(\ref{max})
gives also $f_{\mu\nu}=0$. So, there is a `scalar field' $\psi$,
\[ \psi_{,\mu}=\Phi_\mu \,, \]
which is responsible for the longitudinal polarization \cite{zh}.
The existence of spherically symmetric non-stationary (and non-singular)
solutions is proved in \cite{zh}. 

GR cosmologies with linear expansion  still do exist, e.g.\
\cite{melia}, but require a sort of fine tuning of DE.

\vspace{3.5mm}

\noindent  4. What is the truth and can we find it in physics?
Okay, take a bit more concrete question: should the true
fundamental theory (its existence can explain the usefulness of
phenomenological theories and models) have some clear and
sensible distinction from enumerable false theories? If so, the
true theory, the true description of Nature, has to have no free
parameters; all {\sl fundamental constants} should be
long-lasting, slow-varying parameters of a solution, not of the
theory.\footnote{The string theory (it is not yet a ready
theory; it's not a clear road, just a direction) still has one
free parameter---the string tension. (Nature should keep the
value in every point? In which storage?)} [And I am with those
who think that quantum mechanics (which still needs `classical
crutches' to make predictions; psi-function, amplitude has no
intrinsic, all-sufficient, independent meaning or ontology) is
not the only possible (nor the best) starting point for the way
to a deep insight into the origins of Nature.]

The unique variant of AP is a very promising 5$D$ theory, with
topological charges and quasi-charges; their phenomenology can
look, at some conditions and to a certain extent, like a 4$D$
quantum field theory.

 Despite clear lack of development, the theory can give
  qualitative but important explanations (lepton flavors et cet.)
   as well as predictions
   (which can be falsified in future LHC experiments):

   (1)
  the absence of  spin zero elementary particles
  (there are no SO$_3$-symmetrical quasi-charges);

  (2) there is no room
  for supersymmetry and dark matter (GR
  should be replaced with the forth order 
$R^{\mu\nu}G_{\mu\nu}$-gravity on the brane, with the large extra dimension, plus an additional second order constraint where the Ricci tensor is linearly
coupled with the longitudinal polarization).


\end{document}